\documentclass[floatfix,aps,prd,preprint,eqsecnum,tightenlines,nofootinbib,showpacs]{revtex4}
\usepackage{graphicx,epsf,amsmath}
\usepackage{subfigure}
\usepackage{relsize}
\usepackage{multirow}
\usepackage{array}
\usepackage{verbatim}

\newif\ifdraft
\drafttrue
\newif\ifpreprint

\def\fig#1{Fig.~{\ref{#1}}}

\def\eps{\epsilon}

\def\pol{\varepsilon}
\def\f{\tilde f}

\def\V{{\rm V}}

\def\f{{\!f}}

\def\spa#1.#2{\left\langle#1\,#2\right\rangle}
\def\spb#1.#2{\left[#1\,#2\right]}
\def\tree{{\rm tree}}

\def\Tr{\, {\rm Tr}}

\def\eps{\epsilon}

\def\eqn#1{Eq.~(\ref{#1})}

\def\eqns#1#2{Eqs.~(\ref{#1}) and~(\ref{#2})}

\def\NeqOne{{{\cal N}=1}}

\def\NeqFour{{{\cal N}=4}}
\def\NeqFive{{{\cal N}=5}}
\def\NeqSix{{{\cal N}=6}}
\def\NeqEight{{{\cal N}=8}}

\def\tree{{\rm tree}}

\newbox\charbox
\newbox\slabox
\def\s#1{{      
        \setbox\charbox=\hbox{$#1$}
        \setbox\slabox=\hbox{$/$}
        \dimen\charbox=\ht\slabox
        \advance\dimen\charbox by -\dp\slabox
        \advance\dimen\charbox by -\ht\charbox
        \advance\dimen\charbox by \dp\charbox
        \divide\dimen\charbox by 2
        \raise-\dimen\charbox\hbox to \wd\charbox{\hss/\hss}
        \llap{$#1$} }}

\begin{document}

UCLA/14/TEP/110\hfill $\null\hskip 4cm \null$  \hfill

\title{The Ultraviolet Critical Dimension of Half-Maximal Supergravity
at Three Loops}

\author{Zvi~Bern${}^a$, Scott~Davies${}^a$ and Tristan Dennen${}^b$}
\affiliation{
${}^a$Department of Physics and Astronomy\\
University of California at Los Angeles\\
Los Angeles, CA 90095-1547, USA \\$\null$ \\
${}^b$Niels Bohr International Academy and Discovery Center\\
The Niels Bohr Institute\\
Blegdamsvej 17,\\
DK-2100 Copenhagen
 \O, Denmark
\\
$\null$
\\
$\null$
\\
}

\begin{abstract}
\vskip .5 cm We determine the minimum dimension at which a divergence
first occurs in half-maximal 16-supercharge supergravity at three
loops.  For the four-point amplitudes, we find that the critical
dimension is $D=14/3$ and give the explicit form of the divergence.
We also give the divergence of an additional piece that gives us
access to theories with higher degrees of supersymmetry, in particular
20- and 32-supercharge supergravities.  We give the divergences of
half-maximal supergravity when matter vector multiplets are included
as well.  Explicit forms of various divergences at one and two loops
are also given.
\end{abstract}

\pacs{04.65.+e, 11.15.Bt, 11.25.Db, 12.60.Jv \hspace{1cm}}

\maketitle

\section{Introduction}

The study of ultraviolet divergences in gravity theories has a long
history dating back to the work of 't~Hooft and
Veltman~\cite{tHooftVeltman}.  Today, consensus opinion holds that all
supergravity theories are necessarily ultraviolet divergent at some
finite loop order.  Indeed, recent studies of supergravity symmetries
find that four-dimensional $\NeqFour$, $\NeqFive$ and $\NeqEight$
supergravity theories~\cite{N8SUGRA, N4SUGRA} appear to have valid
counterterms at three, four and seven loops, respectively, suggesting
that these theories will likely diverge at these loop
orders~\cite{DivergencePredictions, BjornssonGreen}.  However, more
recent calculations demonstrate the existence of a new type of
nontrivial multiloop ultraviolet cancellation called ``enhanced
cancellation''~\cite{Enhanced}.  These cancellations are not accounted for in
analyses based on standard-symmetry arguments.  Enhanced
cancellations have been explicitly
exhibited in calculations in four-dimensional pure $\NeqFour$ and
$\NeqFive$ supergravities where the expected divergences at
three and four loops, respectively, are in fact not
present~\cite{threeLoopHalfMax,Enhanced}. These calculations are made
possible by the duality between color and
kinematics~\cite{BCJ,BCJLoop} and advanced integral reductions, as
implemented in {\tt FIRE}~\cite{FIRE}. (See also
Ref.~\cite{TourkineVanhove} for a string-theory understanding of the
finiteness in $\NeqFour$ supergravity.)

Enhanced cancellations are a supergravity phenomenon distinct from
cancellations that control the leading ultraviolet behavior of
(supersymmetric) gauge theory.  Unlike gauge theory, an analysis of
unitarity cuts in supergravity shows that enhanced cancellations
cannot be made manifest diagram by diagram in any covariant local
diagrammatic formalism based on ordinary Feynman
propagators~\cite{Enhanced}. For $\NeqEight$ supergravity, the
diagram-by-diagram power counting of maximal unitarity cuts is
identical to the earlier power-counting analysis of Bj\"{o}rnsson
and Green~\cite{BjornssonGreen}.  They use a first quantized
pure-spinor formalism~\cite{Berkovits} that exposes all conventional
supersymmetry cancellations at the diagrammatic level. Enhanced
cancellations between diagrams would be on top of these.

In principle, one would like to power count the supergravity
integrands directly in order to understand the origin of
cancellations.  However, this is not feasible because the
cancellations occur not only between different diagrams, but in
particular between planar and nonplanar diagrams.  We would need a
global set of variables along with a formulation that exhibits
integrand-level cancellations between diagrams. Unfortunately it is
not known how to choose such variables for nonplanar diagrams.  We
therefore integrate the expressions in order to expose any
potential ultraviolet cancellations.

There has been an attempt to explain the observed cancellations in
$\NeqFour$ supergravity at three loops using standard supergravity
symmetries by assuming the existence of a
non-Lorentz covariant off-shell 16-supercharge harmonic
superspace~\cite{HarmonicSuperspace}.
However, the required superspace predicts continued finiteness at
three loops when matter is added to the theory, which is in
contradiction to subsequent computations~\cite{halfMaxMatter}.  The
finiteness therefore remains unexplained by standard-symmetry
arguments.

What then might be behind enhanced cancellations?  An analysis of
the simpler but analogous case of half-maximal supergravity in $D=5$
at two loops points to the duality between color and kinematics as
being responsible~\cite{halfMax}.  Unfortunately, this analysis is
difficult to generalize to higher loops.  It is therefore
important to gather further data on the structure of ultraviolet
divergences in supergravity theories, as we do in this paper.

Here we add to the available knowledge on the ultraviolet divergence
structure of scattering amplitudes in supergravities by studying the
three-loop half-maximal-supergravity critical dimension, which is
defined as the lowest spacetime dimension where an ultraviolet
divergence occurs.  We give the divergence of half-maximal
supergravity in the critical dimension, as well as the divergence for
an additional piece that allows us alter the field content to study
theories with higher degrees of supersymmetry.  To perform the
computations, we analytically continue the loop momenta to a higher
dimension $D$ while keeping state-count parameters that allow us to
tune the number of physical states.  Different values for these
parameters will give us access to different theories.  Half-maximal
and maximal supergravities, which in four dimensions are the
$\NeqFour$ and $\NeqEight$ theories, respectively, have a natural
definition in higher dimensions in terms of a dimensional continuation
from a ten-dimensional theory.  In fact, the maximal supergravity
four-point integrand is unaltered through at least four loops as one
changes the dimension~\cite{HigherDim}.  $\NeqFive$ supergravity,
however, is a purely four-dimensional theory and is not well-defined
in higher dimensions.  Nevertheless, we can use four-dimensional
values for the state-count parameters and search for additional
cancellations with only the loop momenta analytically continued to
higher dimensions.

As already mentioned, pure half-maximal $\NeqFour$ supergravity is three-loop
finite in four dimensions~\cite{threeLoopHalfMax}.  The next
dimension where there could be a divergence is $D=14/3$, and we find
that indeed half-maximal supergravity diverges here.  We give the
explicit value of the divergence and also collect various
higher-dimensional one- and two-loop divergences.  We give similar
divergences for our additional piece that allows us access to
theories with more supersymmetry.

This paper is organized as follows.  In Sect.~\ref{sec:methods}, we
summarize the methods that we use to obtain our results.  In
Sect.~\ref{sec:oneloop}, we give explicit values of higher-dimensional
divergences at one loop.  Then in Sect.~\ref{sec:twoloop}, we give
divergence results at two loops.  Sect.~\ref{sec:threeloop} contains
the main results of the paper: the divergences of half-maximal
supergravity with and without matter at three loops in $D=14/3$. We
also discuss the three-loop divergences of theories with more
supersymmetry through our additional piece.  In
Sect.~\ref{sec:conclusion}, we give our concluding remarks.

\section{Methods}
\label{sec:methods}

In this section, we summarize the methods that we use to study the
ultraviolet properties of supergravity theories.  These methods have
been previously discussed in detail in Ref.~\cite{halfMaxMatter}, so
here we remain brief.

\subsection{Duality between color and kinematics}

To obtain supergravity loop integrands, we use the duality between
color and kinematics~\cite{BCJ,BCJLoop}.  This allows us to build
supergravity integrands directly from simpler corresponding
gauge-theory integrands.

\begin{figure}
\includegraphics[scale=.7]{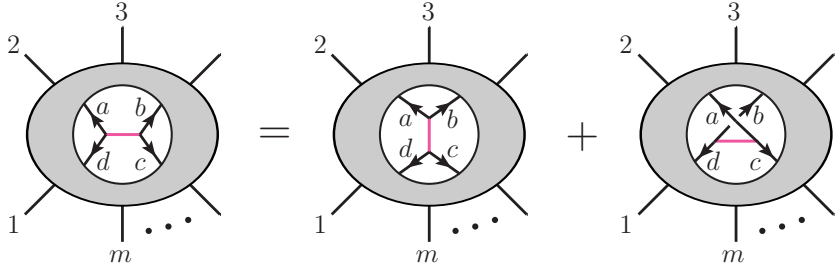}
\caption{The basic Jacobi relation between three diagrams for either
  color or numerator factors.  The basic identity can be embedded into
  a diagram at any loop order. }
\label{fig:GeneralJacobi}
\end{figure}

The duality between color and kinematics is usually expressed in terms
of diagrams with only three-point vertices.  A generic color-dressed
Yang-Mills amplitude at $L$ loops may be written in terms of such
diagrams as
\begin{align}
\mathcal{A}_m^{(L)}=i^Lg^{m-2+2L}\sum_{{\cal S}_m}\sum_j
\int\prod_{l=1}^L\frac{d^Dp_l}{(2\pi)^D}
\frac{1}{S_j}\frac{n_jc_j}{\prod_{\alpha_j}p_{\alpha_j}^2}\,.
\label{eq:YMBCJ}
\end{align}
The sum over $j$ is over all trivalent diagrams; contributions
containing four-point vertices are assigned to these diagrams by
multiplying and dividing by appropriate propagators.  There is also a
sum over the $m!$ permutations of the external legs and a symmetry
factor $S_j$ to remove overcounts due to graph automorphisms.  The
numerator factor $c_j$ is the color factor of graph $j$ obtained by
dressing each three-vertex with a structure constant
$\tilde{f}^{abc}=i\sqrt{2}f^{abc}$.  The numerator factor $n_j$ is the
kinematic numerator containing dependence on momenta, polarizations, spinors
and, if a superspace formalism is used, Grassmann parameters.  In the
denominator, $p_{\alpha_j}^2$ are the propagators of graph $j$.

The color-kinematics duality conjecture states that there exists a
representation of amplitudes in the form of Eq.~\eqref{eq:YMBCJ} such
that the kinematic numerators $n_j$ obey the same algebraic properties
as color factors. In particular, the numerators obey the same Jacobi
relations as the color factors of the same diagrams (see
Fig.~\ref{fig:GeneralJacobi}):
\begin{align}
c_i=c_j-c_k \Rightarrow n_i=n_j-n_k\,.
\label{eq:duality}
\end{align}
The duality has been proven at tree level~\cite{BCJproof}; at loop
level, it remains a conjecture, but we use it only
where duality-satisfying representations have been explicitly constructed.

\begin{figure}
\centering
\subfigure[]{\includegraphics[scale=0.45]{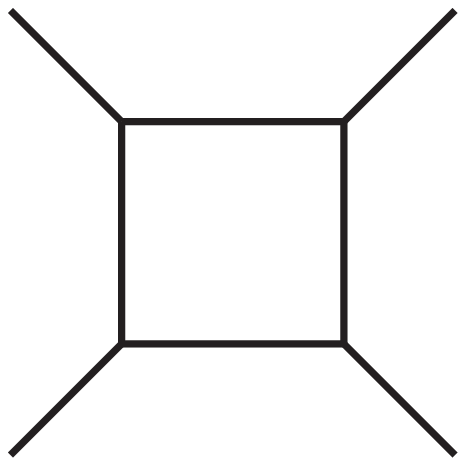}}
  \hskip 2.5 cm
\subfigure[]{\includegraphics[scale=0.45]{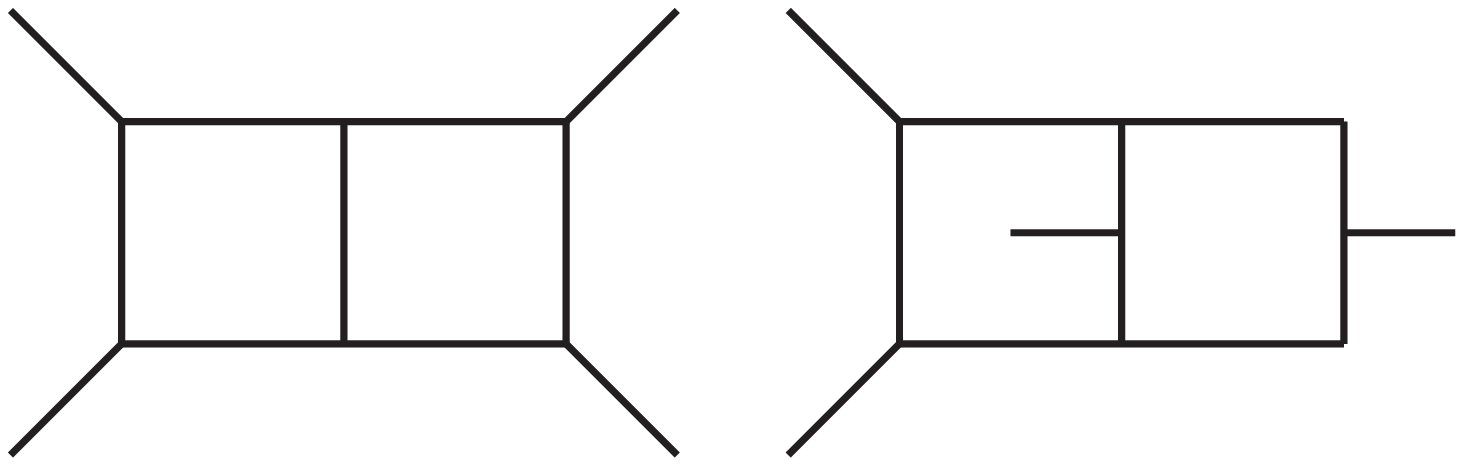}}
\caption[]{The contributing diagrams at (a) one loop and (b) two loops
in the double-copy construction of the four-point amplitude in supergravity 
theories with 16 or more supercharges.} 
\label{fig:OneTwoLoop}
\end{figure}

Once a duality-satisfying representation of a Yang-Mills amplitude is
found, we obtain corresponding gravity amplitudes through the
double-copy procedure by simply replacing the color factor in
Eq.~\eqref{eq:YMBCJ} with a duality-satisfying numerator of a second
gauge theory: 
\begin{equation}
c_j \rightarrow \tilde{n}_j\,.
\end{equation}
After a trivial replacement of the coupling constant, this
gives us the corresponding gravity amplitude,
\begin{align}
\mathcal{M}_m^{(L)}=i^{L+1}\left(\frac{\kappa}{2}\right)^{m-2+2L}
\sum_{S_m}\sum_j\int\prod_{l=1}^L\frac{d^Dp_l}{(2\pi)^D}
\frac{1}{S_j}\frac{n_j\tilde{n}_j}{\prod_{\alpha_j}p_{\alpha_j}^2}\,.
\label{eq:doubleCopy}
\end{align}
The two numerators may belong to different Yang-Mills theories: By
choosing different combinations of Yang-Mills theories, we obtain
amplitudes in different (super)gravity theories.  An important feature
is that only one of the two copies $n_j$ or $\tilde{n}_j$ needs to
satisfy the duality \eqref{eq:duality} for
\eqn{eq:doubleCopy} to be valid~\cite{BCJLoop,GravSquare}.  Furthermore, if it happens that one
of the gauge-theory numerators vanishes for a certain diagram, then we
do not need the corresponding numerator in the other gauge theory.  In
particular, for the one- and two-loop four-point amplitudes of ${\cal
  N} \ge 4$ supergravity, the only required diagrams are those
displayed in \fig{fig:OneTwoLoop}.  All other diagrams have vanishing
numerators in the maximally supersymmetric Yang-Mills component of the
double-copy construction.

\subsection{Construction of supergravity integrands}   

We now consider in more detail the specific constructions required for
this paper.  We start with half-maximal supergravity, which is a
16-supercharge theory.  In four dimensions, it is the $\mathcal{N}=4$
supergravity theory~\cite{N4SUGRA}.  We obtain half-maximal
supergravity by taking one numerator in the double
copy~\eqref{eq:doubleCopy} to be from maximally supersymmetric
Yang-Mills theory and the second from ordinary nonsupersymmetric
Yang-Mills theory:
\begin{align}
Q=16\,\mathrm{sugra}\,\,\,:\,\,\,(Q=16\,\mathrm{sYM})
\otimes(Q=0\,\mathrm{sYM})\,,
\end{align}
where $Q$ counts the number of supercharges, and ``sugra'' and ``sYM''
are shorthands for, respectively, supergravity and super-Yang-Mills.
By $Q=0$ sYM, we mean ordinary nonsupersymmetric Yang-Mills theory.
In $D=4$, $Q=16$ sYM is the $\mathcal{N}=4$ super-Yang-Mills theory.
In this paper we calculate half-maximal supergravity divergences at
one, two and three loops.  In doing so, we leave as arbitrary a
state-counting parameter $D_s$ on the pure Yang-Mills side.  $D_s$ is
the spin-state dimension of the internal gluons, i.e. the internal
gluons have $D_s-2$ spin states.  The loop momenta is still in $D$
dimensions.  Via dimensional reduction then, we are able to include
$(D_s-D)$ real scalars coupled to the nonsupersymmetric Yang-Mills
theory in $D$ dimensions.  In half-maximal supergravity, this
corresponds to including $n_\V = (D_s - D)$ matter vector multiplets
in the theory.

We also calculate divergences of a piece that we call the ``$N_\f$
piece''.  It is obtained by taking a direct product of maximally
supersymmetric Yang-Mills theory with nonsupersymmetric Yang-Mills
theory coupled to non-chiral adjoint fermions, keeping only terms where
each nonsupersymmetric Yang-Mills contribution contains at least one
fermion loop.  We define
\begin{equation}
N_\f = n_\f \Tr[1]\,,
\label{DfDef}
\end{equation}
where $\Tr[1]$ gives the dimensionality of the gamma matrices, and
$n_\f$ is the number of fermion flavors.  The parameter $N_\f$ is thus
the number of fermionic states of one of the gauge theories from which
the gravity theory is composed, not counting color degrees of freedom.
$Q=16$ super-Yang-Mills theory contains 16 states in any dimension, so
the total number of states in the corresponding
supergravity theory is $16 (N_f + D_s-2)$.  We again allow for an
arbitrary number of scalars in the $N_\f$ piece through the parameter
$D_s$.  The $N_\f$ piece is the difference between two supergravity
calculations:
\begin{align}
N_\f\,\mathrm{piece}\,\,\,:\,\,\,\,&(Q=16\,\mathrm{sYM})
 \otimes(Q=0\,\mathrm{sYM}+ N_\f \,\mathrm{fermion\ states}) \notag \\
&\ominus(Q=16\,\mathrm{sYM})\otimes(Q=0\,\mathrm{sYM})\,,
\end{align}
where $(Q=0\,\mathrm{sYM}+ N_\f \,\mathrm{fermion\ states})$
represents pure nonsupersymmetric Yang-Mills theory coupled to non-chiral
adjoint fermions.  While this formula is valid for any states in the
theory, in this paper we will consider only external states
constructed from any state on the $Q=16$ super-Yang-Mills side and
only gluons from the other side.  There is no restriction on the
states circulating in the loops.  This decomposition allows us to
extend half-maximal supergravity results to theories with higher
degrees of supersymmetry by controlling the number of scalars and
non-chiral fermions on the non-maximal super-Yang-Mills side.

For certain choices of the parameters $D_s$ and $N_\f$ and loop
integration dimension $D$, the obtained theories correspond to
well-known supergravity theories.  For example, in
Ref.~\cite{Enhanced}, we promoted $\mathcal{N}=4$ supergravity in four
dimensions to $\mathcal{N}=5$ supergravity in $D=4$ by including the
$N_\f$ piece and choosing $D_s=4$ and $N_\f=2$.  The end result is
equivalent to taking a direct product of $\mathcal{N}=4$ super-Yang-Mills
theory and $\mathcal{N}=1$ super-Yang-Mills theory.  Including both
the half-maximal piece and the $N_\f$ piece for divergences in this
paper will allow us to easily examine cases besides half-maximal
supergravity, most notably maximally supersymmetric supergravity.
Many results in the maximal theory are already known, so it provides
useful checks on our results.  For the three-loop case, we also look
at $\NeqFive$ supergravity by keeping the state counts at their
four-dimensional values.  For generic values of the state-counting
parameters, this construction amounts to analytic continuations of the
well-known supergravity theories.  We note that our construction is
not valid for four-dimensional $\NeqSix$ supergravity.  We can obtain
this from a dimensional reduction of $Q= 24$ or ${\cal N} = 3$
supergravity in $D=6$, but that would require six-dimensional chiral
fermions for the $N_\f$ contributions, which we do not incorporate in
the present calculation.

In our construction we use the color-kinematics duality-satisfying
representations on the maximal $Q=16$ super-Yang-Mills
side~\cite{BCJLoop}.  Since only one copy in Eq.~\eqref{eq:doubleCopy}
needs to satisfy the duality~\eqref{eq:duality}, we are free to use
any other convenient representation on the nonsupersymmetric side of
the double copy.  At four points, the one- and two-loop amplitudes are
especially easy to obtain because the maximal super-Yang-Mills
kinematic numerators are independent of loop momenta~\cite{Camille,halfMax}.
This means that the corresponding supergravity amplitudes are linear
combinations of gauge-theory ones even after integration.  Three loops
is more complicated. At three loops, we use Feynman rules for the
non-maximally supersymmetric side. While this may seem inefficient, it
is a good choice for our purposes. We only need to keep those Feynman
diagrams that contain color factors corresponding to nonvanishing
numerators on the maximally supersymmetric Yang-Mills side.  We also
use Feynman diagrams as a cross check on our one- and two-loop results
below.

\subsection{Extraction of ultraviolet divergences}

After using the double-copy formula (\ref{eq:doubleCopy}) to construct
the integrands of the half-maximal and $N_\f$ pieces, we must integrate
to extract the ultraviolet divergence.  Our procedure is described in
detail in Ref.~\cite{halfMaxMatter}; here we briefly summarize it.

Since we are only interested in the ultraviolet behavior, we series
expand power-divergent integrals in large internal loop momenta, or
equivalently small external momenta~\cite{SeriesAndSubs}.  In this
expansion, only the logarithmically divergent integrals contribute to
the ultraviolet divergence, so we drop all other pieces.  After removing
subdivergences, the
divergences of logarithmically divergent integrals are proportional to
numbers with no dependence on external momenta.  We may therefore set
the external momenta in the resulting integrals to zero, converting
the integrals to vacuum integrals that are much easier to evaluate.

The resulting vacuum integrals are badly infrared divergent, so we
need regulators that separate these singularities from the ultraviolet
ones that we are interested in obtaining. For the ultraviolet
divergences we use dimensional reduction~\cite{DimRed}. A particularly
good choice of infrared regulator is to assign all propagators a
uniform mass at the start of the calculation~\cite{UniformMass}.  In
many cases, this uniformity allows for cancellations at lower loop
orders to feed into the calculation, eliminating the need to subtract
subdivergences integral by integral~\cite{SeriesAndSubs}.  For the
three-loop cases in $D=14/3$, none of the individual integrals
actually have subdivergences, so subtractions are not required.
However, for the two-loop cases in six dimensions examined here,
subdivergence subtractions are required even with a uniform mass due
to quadratic subdivergences, which apparently break the feed-through
of lower-loop cancellations.  An advantage of explicitly subtracting
subdivergences is that it allows us to avoid potential subtleties
associated with evanescent operators that might occur when using
counterterms to remove lower-loop subdivergences~\cite{Jones}.

Once we have the logarithmically-divergent, mass-regulated vacuum
integrals, we reduce to a basis of master integrals using {\tt
  FIRE}~\cite{FIRE}, which implements integration-by-parts identities
using the Laporta algorithm~\cite{Laporta}.  In $D=4$ the master
integrals are available in the literature since they correspond to
ones used in the calculations of the three- and four-loop QCD $\beta$
functions~\cite{QCDbeta}. We also perform the integrals explicitly
using Mellin-Barnes techniques and resummation of
residues~\cite{IntegralMethods}.  The three-loop integrals in $D=14/3$
required for our computation are evaluated this way.

\section{One-loop divergences}
\label{sec:oneloop}

At one loop, all pure supergravities are finite in
$D=4$~\cite{OneLoopFinite}. They are also finite in $D=6$ because the
required $R^3$ counterterm violates supersymmetry identities of a
four-dimensional subspace~\cite{TwoLoopGrisaru}.  Consistent with
this, we find that both the half-maximal piece and the $N_\f$ piece
are separately finite in these dimensions.  With matter, however,
half-maximal supergravity is divergent in both $D=4$ and $D=6$.
Refs.~\cite{OneLoopN4Matter,halfMaxMatter} contain the explicit forms
of the divergences for external matter in four dimensions;
Ref.~\cite{halfMaxMatter} contains divergences for external matter in
six dimensions.

In this section we will focus on $D=8$, which is the one-loop critical
dimension where divergences first occur when there are no matter
multiplets.  We first review $D=8$ one-loop results in half-maximal
supergravity and give new results for the $N_\f$ piece.  The
divergence for half-maximal supergravity in $D=8$ was first given in
Ref.~\cite{DunbarOneLoop}.  Divergences involving external matter
states in eight dimensions can be found in Ref.~\cite{halfMaxMatter}.

\subsection{Eight dimensions}

In eight dimensions, the divergence for pure half-maximal supergravity
with internal matter is~\cite{DunbarOneLoop,halfMax}
\begin{align}
\mathcal{M}^{(1)}_{Q=16}(1,2,3,4)\Big|_{D=8\, \rm div.} = &
-\frac{1}{\eps}  \frac{1} {(4 \pi)^4}  
\left(\frac{\kappa}{2}\right)^4stA^{\rm tree}_{Q = 16}(1,2,3,4) \notag \\
&\times\left[ \frac{238+D_s}{360}(F_1F_2F_3F_4) + \frac{D_s-50}{288}(F_1 F_2)(F_3 F_4)
\right]\notag\\
& \null\hskip 3 cm  +{\rm cyclic(2,3,4)} \,,
\label{GravityOneLoopD8}
\end{align} 
where $\eps = (8-D)/2$ is the usual dimensional regularization
parameter, 
 $A^{\rm tree}_{Q = 16}(1,2,3,4)$ is the maximal super-Yang-Mills
four-point tree amplitude, and ${\rm cyclic(2,3,4)}$ represents the cyclic
permutations over legs 2, 3 and 4. We use the Mandelstam invariants,
\begin{equation}
s = (k_1+k_2)^2\,, \hskip 1 cm t = (k_2 + k_3)^2\,, \hskip 1 cm u = (k_1 + k_3)^2\,.
\end{equation}
Here
\begin{align}
(F_i F_j) \equiv F_i^{\mu\nu} F_{j\mu\nu}\,,\hskip 1 cm 
(F_{i}F_{j}F_{k}F_{l})\equiv F_i\,^{\mu\nu}F_{j\nu\rho}F_k\,^{\rho\sigma}F_{l\sigma\mu}\,,
 \end{align}
where the linearized polarization field strength of each leg $j$ is
\begin{equation}
F_j^{\mu\nu} \equiv i( k_j^\mu \pol_j^\nu - k_j^\nu \pol_j^\mu) \,.
\label{FtoAmp}
\end{equation}
The divergence can also be generated by the operator~\cite{DunbarOneLoop,halfMax},
\begin{align}
\mathcal{O}&=-\frac{1}{\epsilon}\frac{1}{(4\pi)^4}\frac{1}{11520}
[(-126+3D_s)T_1+(1968-24D_s)T_2+(-252+6D_s)T_3 \notag \\
&\hspace{3.5cm} + (8-4D_s)T_4+3840T_5-1920T_6+(-3776-32D_s) T_7 ] \,,
\label{DunbarR4}
\end{align}
where
\begin{align}
T_1&=(R_{\mu\nu\rho\sigma}R^{\mu\nu\rho\sigma})^2, \notag \\
T_2&=R_{\mu\nu\rho\sigma}R^{\mu\nu\rho}_{\hphantom{\mu\nu\rho}\lambda}R_{\gamma\delta\kappa}^{\hphantom{\gamma\delta\kappa}\sigma}R^{\gamma\delta\kappa\lambda}, \notag \\
T_3&=R_{\mu\nu\rho\sigma}R^{\mu\nu}_{\hphantom{\mu\nu}\lambda\gamma}R^{\lambda\gamma}_{\hphantom{\lambda\gamma}\delta\kappa}R^{\rho\sigma\delta\kappa}, \notag \\
T_4&=R_{\mu\nu\rho\sigma}R^{\mu\nu}_{\hphantom{\mu\nu}\lambda\gamma}R^{\rho\lambda}_{\hphantom{\rho\lambda}\delta\kappa}R^{\sigma\gamma\delta\kappa}, \notag \\
T_5&=R_{\mu\nu\rho\sigma}R^{\mu\nu}_{\hphantom{\mu\nu}\lambda\gamma}R^{\rho\hphantom{\delta}\lambda}_{\hphantom{\rho}\delta\hphantom{\lambda}\kappa}R^{\sigma\delta\gamma\kappa}, \notag \\
T_6&=R_{\mu\nu\rho\sigma}R^{\mu\hphantom{\lambda}\rho}_{\hphantom{\mu}\lambda\hphantom{\rho}\gamma}R^{\lambda\hphantom{\delta}\gamma}_{\hphantom{\lambda}\delta\hphantom{\gamma}\kappa}R^{\nu\delta\sigma\kappa}, \notag \\
T_7&=R_{\mu\nu\rho\sigma}R^{\mu\hphantom{\lambda}\rho}_{\hphantom{\mu}\lambda\hphantom{\rho}\gamma}R^{\lambda\hphantom{\delta}\nu}_{\hphantom{\lambda}\delta\hphantom{\nu}\kappa}R^{\gamma\delta\sigma\kappa} \,.
\end{align}
We note that on shell the combination
\begin{equation}
-\frac{T_1}{16}+T_2-\frac{T_3}{8}-T_4+2T_5-T_6+2T_7\,,
\label{onShellVanish}
\end{equation}
is a total derivative~\cite{DunbarOneLoop}, so there is freedom in
rewriting Eq.~\eqref{DunbarR4}.  For four-dimensional external states
in an MHV-helicity configuration, we can write the divergence in terms
of four-dimensional spinor helicity \cite{halfMax}:
\begin{align}
\mathcal{M}_{Q=16}^{(1)}(1^-,2^-,3^+,4^+)\Big|_{D=8\,\mathrm{div.}}=\frac{i}{\epsilon}\frac{1}{(4\pi)^4}\left(\frac{\kappa}{2}\right)^4\frac{58+D_s}{180}\langle 1\,2\rangle^4[3\,4]^4\,.
\end{align}

For the $N_\f$ piece, we have
\begin{align}
\mathcal{M}^{(1)}_{N_\f}(1,2,3,4)\Big|_{D=8\, \rm div.} =&
-\frac{1}{\eps}  \frac{1} {(4 \pi)^4} N_\f
 \left(\frac{\kappa}{2}\right)^4stA^{\rm tree}_{Q = 16}(1,2,3,4) \notag \\
&\times\left[ \frac{7}{180}(F_1F_2F_3F_4) - \frac{1}{72}(F_1 F_2)(F_3 F_4)
\right] +{\rm cyclic(2,3,4)} \,.
\label{GravityOneLoopD8ferm}
\end{align} 
This can also be generated by the operator,
\begin{align}
\mathcal{O}&=\frac{1}{\epsilon}\frac{1}{(4\pi)^4} N_\f \frac{1}{23040}
[21T_1-288T_2+42T_3-8T_4-480T_5+240T_6+416T_7 ] \,.
\label{R4ferm}
\end{align}
For four-dimensional external states in an MHV-helicity configuration, we have
\begin{align}
\mathcal{M}^{(1)}_{N_\f}(1^-,2^-,3^+,4^+)\Big|_{D=8\,\,\mathrm{div.}}=\frac{i}{\epsilon}\frac{1}{(4\pi)^4}\left(\frac{\kappa}{2}\right)^4 N_\f\frac{11}{720}\langle 1\,2\rangle^4[3\,4]^4\,.
\end{align}

As a check on these results, we obtain the result for maximal
supergravity by setting $D_s=10$ and $N_\f=8$.  This gives
\begin{align}
\mathcal{M}^{(1)}_{Q=32}(1,2,3,4)\Big|_{D=8\, \rm div.} &=
-\frac{1}{\eps}  \frac{1} {(4 \pi)^4}  \left(\frac{\kappa}{2}\right)^4stA^{\rm tree}_{Q = 16}(1,2,3,4) \notag \\
&\hspace{1cm}\times\left[(F_1F_2F_3F_4) -\frac{1}{4}(F_1 F_2)(F_3 F_4)
\right] +{\rm cyclic(2,3,4)} \notag \\
&=-\frac{i}{\epsilon}\frac{1} {(4 \pi)^4}  \left(\frac{\kappa}{2}\right)^4\frac{1}{2}\left[stA^{\rm tree}_{Q = 16}(1,2,3,4)\right]^2\,,
\label{GravityOneLoopD8N8}
\end{align} 
in agreement with the known result~\cite{NEq8OneLoop}.  The gravity
operator corresponding to this divergence is~\cite{DunbarOneLoop}
\begin{align}
\mathcal{O}&=\frac{1}{\epsilon}\frac{1}{(4\pi)^4}\frac{1}{64}
[T_1-16T_2+2T_3-32T_5+16T_6+32T_7 ] \,.
\end{align}
For the MHV four-graviton configuration in terms of four-dimensional spinor helicity, this is 
\begin{align}
\mathcal{M}^{(1)}_{Q=32}(1^-,2^-,3^+,4^+)\Big|_{D=8\,\,\mathrm{div.}}
=\frac{i}{\epsilon}\frac{1}{(4\pi)^4}\left(\frac{\kappa}{2}\right)^4\frac{1}{2}\langle 1\,2\rangle^4[3\,4]^4\,.
\end{align}

\section{Two-loop divergences}
\label{sec:twoloop}

It has been known since the early days of supergravity that all pure
theories without matter are two-loop finite in $D=4$ because no valid
counterterms exist~\cite{TwoLoopGrisaru,TwoLoopFinite}.  In five
dimensions, however, an apparently valid $R^4$ counterterm exists for
half-maximal supergravity under supersymmetry and duality-symmetry
constraints~\cite{VanishingVolume,HarmonicSuperspace,halfMaxMatter}.
Nevertheless, a string-theory calculation~\cite{TourkineVanhove} and
an explicit field-theory calculation~\cite{halfMax} demonstrate that
half-maximal supergravity is in fact ultraviolet finite at two loops
in $D=5$. The cancellations are not visible diagram by diagram in any
covariant local diagrammatic formalism, so this case is an example of
enhanced cancellations~\cite{Enhanced}.  Interestingly, the
cancellations can be instead understood by relating the supergravity
amplitudes to Yang-Mills amplitudes involving color tensors that are
forbidden to appear in divergences~\cite{halfMax}.  A key point in
this analysis is that it captures cancellations between diagrams,
particularly between the planar and nonplanar sectors.

The two-loop critical dimension for pure half-maximal supergravity is
$D_c=6$.  The explicit divergence including internal matter vector
multiplets was given in Ref.~\cite{halfMax}.  In this section we
review this result and give the new result of the divergence for the
$N_\f$ piece.  With external matter states, half-maximal supergravity
diverges in $D=4$, 5 and 6; explicit forms can be found in
Ref.~\cite{halfMaxMatter}.

\subsection{Six dimensions}

In $D=6$, the divergence in half-maximal
supergravity with internal-matter multiplets is~\cite{halfMax}
\begin{align}
\mathcal{M}^{(2)}_{Q=16}(1,2,3,4) \Bigr|_{D=6\, \rm div.}  \hspace{-.3cm} = &
\; \frac{1}{(4\pi)^6}\left(\frac{\kappa}{2}\right)^6\!
  st A_{Q=16}^\tree(1,2,3,4)\notag \\
& \!\times \!
\bigg\{
\biggl(\frac{(D_s-6)(26-D_s)}{576\epsilon^2} +
            \frac{19D_s-734}{864 \epsilon}\biggr)  \label{NoNfD6TwoLoop}\\
&\hspace{2cm} \times 
\bigg[ s\, (F_1 F_2)(F_3 F_4) + t\, (F_1 F_4)(F_2 F_3) + u\, (F_1 F_3)(F_2 F_4) 
\bigg]\notag \\
&\hspace{.2cm}+\frac{48D_s-1248}{864\epsilon}\bigg[u \, (F_1 F_2 F_3 F_4)
  + t \, (F_1 F_3 F_4 F_2)+s \, (F_1 F_4 F_2 F_3)\bigg]\bigg\}\,. \notag
\end{align}
The number of vector matter multiplets
is given by $n_\V = D_s - 6$.  The result for pure half-maximal
supergravity is obtained by setting $D_s = 6$, and we find that
even with no matter multiplets, the
theory still diverges.  
The $1/\eps^2$ divergence vanishes in this case because there are
no one-loop subdivergences in the pure theory.

We can simplify the expressions a bit by restricting the external
states to four-dimensional helicity states. For the MHV configuration,
this gives the divergence~\cite{halfMax},
\begin{align}
\mathcal{M}^{(2)}_{Q=16}(1^-,2^-,3^+,4^+)\Big|_{D=6\,\,\mathrm{div.}} = & 
-\frac{i}{(4\pi)^6} \left(\frac{\kappa}{2}\right)^6
s \langle 1\,2\rangle^4[3\,4]^4 \notag \\
& \times
\left(\frac{(D_s-6)(26-D_s)}{576\epsilon^2}
 +\frac{19D_s-734}{864\epsilon}\right) .
\end{align}

The $N_\f$ piece also diverges in six dimensions.  Its form is
\begin{align}
\mathcal{M}^{(2)}_{N_\f}(1,2,3,4) \Bigr|_{D=6\, \rm div.}  \hspace{-.4cm}&=
\!\frac{1}{(4\pi)^6}\left(\frac{\kappa}{2}\right)^6\! N_\f st A_{Q=16}^\tree(1,2,3,4)\bigg\{
\biggl(\frac{6-D_s}{288\epsilon^2} +
            \frac{21D_s+62}{3456 \epsilon}\biggr) \notag \\
&\hspace{2cm} \times 
\bigg[ s\, (F_1 F_2)(F_3 F_4) + t\, (F_1 F_4)(F_2 F_3) + u\, (F_1 F_3)(F_2 F_4) 
\bigg]\notag \\
&\hspace{.8cm}+\frac{3D_s-14}{144\epsilon}\bigg[u \, (F_1 F_2 F_3 F_4)
  + t \, (F_1 F_3 F_4 F_2)+s \, (F_1 F_4 F_2 F_3)\bigg]\bigg\}\,.
\label{NfD6TwoLoop}
\end{align}
For the MHV external helicity configuration, we find
\begin{align}
\mathcal{M}^{(2)}_{N_\f}(1^-,2^-,3^+,.4^+)\Big|_{D=6\,\,\mathrm{div.}}
=\frac{i}{(4\pi)^6}\left(\frac{\kappa}{2}\right)^6 N_\f
s\langle 1\,2\rangle^4[3\,4]^4
\left(\frac{D_s-6}{288\epsilon^2}-\frac{21D_s+62}{3456\epsilon}\right).
\end{align}
As noted earlier, different choices of $N_\f$ and $D_s$ do not
necessarily correspond to supersymmetric theories.  However, the choice $D_s=10$ and
$N_\f=8$ does correspond to maximal supergravity obtained by
dimensional reduction of $\NeqOne$, $D=10$ supergravity to $D=6$.
With this choice, we find that by adding together the contributions in
\eqns{NoNfD6TwoLoop}{NfD6TwoLoop}, the total divergence vanishes.  This
matches a previous calculation that shows that at two loops, maximal
supergravity is ultraviolet finite for $D<7$~\cite{NEq8TwoLoops}.

\section{Three-loop divergences}
\label{sec:threeloop}

\begin{figure}
\includegraphics[scale=.5]{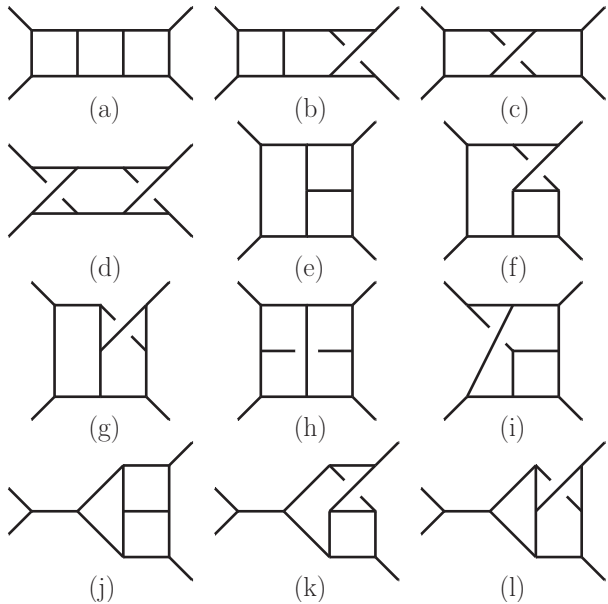}
\caption{The contributing diagrams to the three-loop four-point
amplitudes in theories with 16 or more supercharges.}
\label{fig:TwelveGraphs}
\end{figure}

In Ref.~\cite{threeLoopHalfMax}, the three-loop four-point amplitudes
of pure half-maximal supergravity in $D=4$ (usually called $\NeqFour$
supergravity) were shown to be ultraviolet finite.  The contributing
diagrams are shown in \fig{fig:TwelveGraphs}. The finiteness holds
despite the existence of an apparently valid counterterm under
supersymmetry and duality-symmetry
constraints~\cite{VanishingVolume,HarmonicSuperspace,halfMaxMatter}.
The maximal-cut analysis of Ref.~\cite{Enhanced} shows that
individual diagrams necessarily diverge, so the 
finiteness is a prime example of enhanced cancellations~\cite{Enhanced}.

An interesting question is to identify the critical dimension of the
three-loop amplitudes corresponding to the lowest dimension with
ultraviolet divergences.  The next possible dimension above $D=4$
where divergences can occur for this amplitude is $D=14/3$.  As we now
show, this is indeed the critical dimension for pure half-maximal
supergravity at three loops.  We provide the explicit form of the
divergence, as well as the form of the divergence of the $N_\f$ piece.
We also provide results in $D=14/3$ for half-maximal supergravity
coupled to matter.  Results including matter in $D=4$ can be found in
Ref.~\cite{halfMaxMatter}.

\subsection{$D=14/3$ dimensions with no matter}

Following the steps described in Sect.~\ref{sec:methods}, we find
that the divergence for pure half-maximal supergravity in $D=14/3$ is
\begin{align}
\mathcal{M}^{(3)}_{Q=16}(1,2,3,4)\Big|_{D=14/3\,\,\mathrm{div.}}&=
-\frac{1}{(4\pi)^7}
\left(\frac{\kappa}{2}\right)^8\frac{9(10056-2546 D_s + 99 D_s^2)}
        {61600\epsilon} \notag \\
&\hspace{3cm}\null \times\Gamma^3(\tfrac{1}{3})
 s t A_{Q=16}^{\mathrm{tree}}(1,2,3,4) \, \mathcal{P}\,,
\label{eq:halfMaxDivergence}
\end{align}
where
\begin{align}
\mathcal{P} =&\sum_{{\cal S}_4}
\left(\frac{1}{64} s t u \, \pol_1\cdot\pol_2\pol_3\cdot\pol_4
+\frac{1}{8} t \, \pol_1\cdot\pol_2k_1\cdot\pol_3 (-3 u \, k_1\cdot\pol_4
+(s + 3 t )k_2\cdot\pol_4)\right. \notag \\
&\hspace{1.4cm} \left.\null 
+\frac{1}{12} t\, k_2\cdot\pol_1 k_3\cdot\pol_2(- 4 k_1\cdot\pol_3k_1\cdot\pol_4
- 3 k_2\cdot\pol_3 k_1\cdot\pol_4 + k_2\cdot\pol_3 k_2\cdot\pol_4)\right),
\end{align}
and the sum is over all 24 permutations of the external legs. 
For the $N_\f$ piece, we find
\begin{align}
\mathcal{M}^{(3)}_{N_\f}(1,2,3,4)\Big|_{D=14/3\,\,\mathrm{div.}}= &
-\frac{1}{(4\pi)^7}\left(\frac{\kappa}{2}\right)^8
\frac{3(32136-7738D_s+535D_s^2)}{246400\epsilon}\notag\\
& \hskip .6 cm \null\times
N_\f \, \Gamma^3(\tfrac{1}{3}) 
  s t A_{Q=16}^{\mathrm{tree}}(1,2,3,4)\, \mathcal{P}\,.
\label{eq:nfDivergence}
\end{align}
It is interesting that the potential term proportional to $N_\f^2$
vanishes.

By setting $D_s=10$ and $N_\f=8$, we obtain maximally supersymmetric
supergravity continued to $D=14/3$ dimensions.  With this choice of
parameters, by adding together Eqs.~\eqref{eq:halfMaxDivergence} and
\eqref{eq:nfDivergence}, we find that the divergence cancels, in
agreement with the known finiteness of the maximally supersymmetric
theory at three loops for $D<6$~\cite{NEq8ThreeLoops}.  It is also
interesting to look at the analytic continuation of $\mathcal{N}=5$
theory using four-dimensional state counts.  Setting $D_s=4$ and
$N_\f=2$, we find that the amplitude is still divergent, so there are
no additional cancellations beyond the pure half-maximal theory.
It is interesting to note that the sum of the two divergences in
\eqns{eq:halfMaxDivergence}{eq:nfDivergence} vanishes when we take
$D_s=6$ and $N_\f=4$.  Presumably this cancellation is not accidental.
These contributions give part of the result for four-dimensional
$\mathcal{N}=6$ supergravity analytically continued to $D=14/3$; the
pieces arising from dimensional reduction of the chirality projector
on fermions on the $N_\f$ side of the double copy is missing from our
construction.  It would be interesting to obtain this piece as well to
see if $\mathcal{N}=6$ supergravity with four-dimensional states is
three-loop finite when the loop momenta are analytically continued to
$D=14/3$.

It is also interesting to look at the divergences when the external states
are restricted to live in a four-dimensional subspace where standard
helicity states can be used.  Focusing on the gauge-invariant factor
$\mathcal{P}$, we write helicity terms for the nonsupersymmetric side
of the double copy using four-dimensional spinor helicity:
\begin{align}
&\mathcal{P}_{--++}=0\,, \notag \\
&\mathcal{P}_{-+++}=-\frac{1}{8} s^2 t^2
\frac{[2\,4]^2}{[1\,2]\langle 2\,3\rangle\langle 3\,4\rangle[4\,1]}\,, \notag \\
&\mathcal{P}_{++++}=\frac{5}{4}s t u\frac{[1\,2][3\,4]}{\langle 1\,2\rangle\langle 3\,4\rangle}\,.
\end{align}
where the subscripts  on $\mathcal{P}$ indicate the helicity choices
in an all-outgoing convention.
The vanishing of the MHV configuration is consistent with the fact
that the pure Yang-Mills counterterm is an $F^3$ operator at three
loops in $D=14/3$: $F^3$ cannot generate the MHV configuration at four
points.

\subsection{$D=14/3$ dimensions with matter}

Half-maximal supergravity at three loops in $D=14/3$ coupled to matter 
also diverges as one might expect. The divergences of this theory
in $D=4$ were already discussed in Ref.~\cite{halfMaxMatter}.
For the amplitude with two external graviton multiplets and two
external vector multiplets, we find
\begin{align}
&\mathcal{M}^{(3)}_{Q=16}(1_{\rm{H}},2_{\rm{H}},3_{\rm{V}},4_{\rm{V}})
 \Big|_{D=14/3\,\,\mathrm{div.}} \notag \\
&\hspace{2.5cm}=\null   \frac{1}{(4\pi)^7}\left(\frac{\kappa}{2}\right)^8
\frac{1}{985600\epsilon}\Gamma^3(\tfrac{1}{3}) 
s t A_{Q=16}^{\mathrm{tree}}(1,2,3,4)  \\
&\hspace{4.1cm}\times
\left(5(693D_s^2-67922D_s+443136)\, s^2 \, (2k_1\cdot\varepsilon_2k_2\cdot\varepsilon_1
-s\, \varepsilon_1\cdot\varepsilon_2 )\right. \notag \\
&\hspace{4.5cm}-48(693D_s^2+1678D_s-74688)\, t u \,
k_1\cdot\varepsilon_2k_2\cdot\varepsilon_1  \notag \\
&\hspace{4.5cm}\null
+12(1881D_s^2+3626D_s-195816)\, s t u \, \varepsilon_1\cdot\varepsilon_2 \notag \\
&\hspace{4.4cm}
\null 
+1080(11D_s^2+6D_s-1032) \, s  \notag \\
& \hskip 6 cm \null \left.
\times 
(  s\, k_3\cdot\varepsilon_1k_3\cdot\varepsilon_2 
 - t\, k_1\cdot\varepsilon_2k_3\cdot\varepsilon_1 
 - u\, k_2\cdot\varepsilon_1k_3\cdot\varepsilon_2 )\right)\,, \notag
\end{align}
where the subscript ``$\mathrm{H}$'' indicates a particle belonging to
a graviton multiplet and ``$\mathrm{V}$'' indicates a particle belonging
to a matter vector multiplet.  For four external matter particles
belonging to the same multiplet, we find the divergence,
\begin{align}
\mathcal{M}^{(3)}_{Q=16}(1_{\rm{V}},2_{\rm{V}},3_{\rm{V}},4_{\rm{V}})
 \Big|_{D=14/3\,\,\mathrm{div.}}= &
-\frac{1}{(4\pi)^7}\left(\frac{\kappa}{2}\right)^8
\Gamma^3(\tfrac{1}{3})
    s^2 t^2 u \, A_{Q=16}^{\mathrm{tree}}(1,2,3,4)\, \notag \\
& \hskip 1 cm \null \times
\frac{3}{123200\epsilon}(8811 D_s^2+104656 D_s-1569516) \,.
\end{align}
Finally for external matter from different multiplets, we have
\begin{align}
&\mathcal{M}^{(3)}_{Q=16}(1_{\rm{V_1}},2_{\rm{V_1}},3_{\rm{V_2}},4_{\rm{V_2}})
\Big|_{D=14/3\,\,\mathrm{div.}} \notag \\
&\hspace{1cm}=
\frac{1}{(4\pi)^7}\left(\frac{\kappa}{2}\right)^8
\frac{1}{123200\epsilon}\Gamma^3(\tfrac{1}{3})
 s^2t A_{Q=16}^{\mathrm{tree}}(1,2,3,4)  \\
&\hspace{1.4cm}\times\left(10(495D_s^2-28630D_s+168576)\, s^2
    -3(4587D_s^2-60548D_s+38748)\, t u\right), \notag
\end{align}
where the ``$\mathrm{V}_1$'' and ``$\mathrm{V}_2$'' subscripts
indicate that the particles belong to different matter multiplets.

\section{Conclusions}
\label{sec:conclusion}

Recently a new type of ultraviolet cancellation in multiloop
supergravity amplitudes called enhanced cancellation~\cite{Enhanced}
was uncovered.  The defining characteristic of these cancellations is
that they cannot be made manifest in a covariant local diagrammatic
formalism.  This makes the cancellations nontrivial to study.  At
present there is no explanation for these cancellations based on
supersymmetry and the standard symmetries of
supergravity~\cite{VanishingVolume, HarmonicSuperspace,
  halfMaxMatter}.  There is some evidence at one and two loops that
the duality between color and kinematics~\cite{BCJ,BCJLoop} might be
responsible for the cancellations~\cite{halfMax}.  However, it is
nontrivial to extend this analysis to higher loops.  It therefore is
important to collect as much data as possible on the ultraviolet
structure of supergravity theories.

In this paper we provided new information on the divergence structure
of supergravity theories by analytically continuing the amplitudes to
higher dimensions.  In particular, we analytically continued
half-maximal supergravity to $D=14/3$, corresponding to the lowest
dimension where the three-loop four-point amplitude could
diverge. Indeed, we found a divergence and presented its explicit
form.  We also provided the divergences in this number of dimensions with the
addition of matter multiplets.  Another interesting case that we
studied in $D=14/3$ is the analytic continuation of the three-loop
four-point amplitude of $\NeqFive$ supergravity to this dimension,
keeping the parameters that control the number of states as free
parameters.  Again we found no sensible solution that makes the
amplitude finite.  In any case, the existence of enhanced cancellations
in supergravity theories shows that we have much more
to learn about their ultraviolet structure.

\subsection*{Acknowledgments}
We thank John Joseph Carrasco, Henrik Johansson and Radu Roiban for
discussions.  This work was supported by the Department of Energy
under Award Number DE-{S}C0009937.  We also thank the Danish Council
for Independent Research. We gratefully acknowledge Mani Bhaumik for
his generous support.  We also thank Academic Technology Services at
UCLA for computer support.


\end{document}